\documentclass[twocolumn,nofootinbib,preprintnumbers]{revtex4}

\usepackage{amsmath}
\usepackage{graphicx}

\newcommand{\nn}{\nonumber}

\newcommand{\df}{\mathrm{d}}

\begin{document}

\title{Scale setting and resummation of logarithms in $pp \to V + {\rm jets}$}

\author{Christian W.~Bauer}
\author{Bj{\"o}rn O.~Lange}
\affiliation{Ernest Orlando Lawrence Berkeley National Laboratory, University of California, Berkeley, CA 94720}

\begin{abstract}

The production of vector bosons in association with jets contains at least two unrelated scales. The first is the mass of the vector boson $m_V$ and the second is the hard interaction scale giving rise to large transverse momenta of the produced jets. The production cross sections depend logarithmically on the ratio of these scales, which can lead to a poor convergence in fixed order perturbation theory. We illustrate how to resum all leading logarithmic terms using effective theory methods, and show that they can be resummed by a simple choice of the factorization scale. Implementing this scale choice we show that the large discrepancies between next-to-leading calculations and leading order calculations using more traditional choices of scales disappear.

\end{abstract}

\maketitle

The production of $W$ and $Z$ bosons in association with jets provides one of the most important backgrounds to many searches of physics beyond the standard model (BSM). This is due to the fact that expected models of BSM physics give rise to events containing missing energy (from a potential dark matter candidate), jets (from cascade decays of new strongly interacting particles) and possibly leptons. Since the $W$ and $Z$ bosons can decay to neutrinos and/or leptons, they give rise to events with the same signatures as the BSM signal one is searching for.  To distinguish between the expected new physics and the Standard Model (SM) background one has to study the different kinematical structures of the events. For example, the background preferably  produces jets with small transverse momentum relative to the beam axis, while the cascade decays tend to give rise to much harder jets. However, in isolating potential BSM signature events one often imposes strict kinematic constraints to which the background is also subject to. Thus, a detailed understanding of the differential background spectra is required in order to successfully separate possible BSM signatures from the SM background. 

Much work has gone into detailed calculations of differential distributions for $V + {\rm jets}$, where $V$ denotes either a $W$ or a $Z$ boson. Leading order (LO) calculations~\cite{Alpgen,Madevent,Amegic,Comphep,Grappa,HELAC} are available for up to $6$ jets in the final state, while full next-to-leading order (NLO) calculations are only available for one or two jets in the final state~\cite{Ellis:1981hk,Arnold:1988dp,Giele:1993dj,Campbell:2002tg}. Recently, first NLO calculations for  $V + 3$ jets have become available. In~\cite{Ellis:2009zw} the dominant partonic channels have been calculated at leading order in $1/N_C$, and the full calculation at leading order in $1/N_C$ has followed shortly afterwards~\cite{NLOWjjjBlackhat}. The subdominant $1/N_C$ corrections have been argued to be numerically small, and first results including all $1/N_C$ effects have been presented recently~\cite{Loopfesttalk}. Depending on the choice of scales, the total cross sections at NLO can differ by up to a factor of two from the LO results, while differential spectra can vary even more strongly, especially in the important regime of high $p_T$ jets.

\begin{figure}[t]
\centering
\includegraphics[width=1.0\columnwidth]{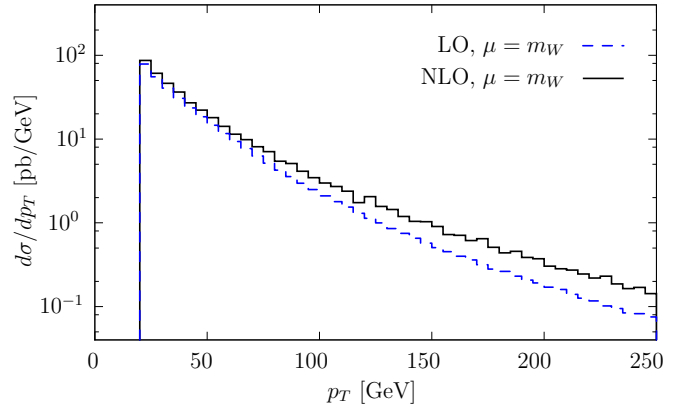}
\caption{$p_T$ distribution for the jet in the $pp\to W^++j$ process, both at leading order (dashed) and next-to-leading order (solid).}
\label{fig:Wj_pTspec}
\end{figure}

It is well known that perturbative corrections generically contain logarithmic terms that depend on ratios of scales in the problem. For the inclusive production of a vector boson, the only relevant scale is the mass of the vector boson $m_{V}$. However, the kinematics of the process (and cuts on these kinematic variables) can give rise to scales that differ significantly from the mass of the produced vector boson. For example, for $V+j$ final states, the transverse momentum $p_T$ of the jet can be far larger than $m_{V}$, and large logarithmic terms of the ratio $p_{T}/m_V$ can potentially spoil the convergence of fixed order perturbation theory. A simple way to illustrate the size of these logarithmic terms is to compare the $p_T$ spectrum at LO and NLO. Since the NLO calculations contain the logarithms of the form $\alpha_s \log^2 p_T/m_V$ and $\alpha_s \log p_T/m_V$, one expects the difference between LO and NLO calculations to get worse as the ratio $p_T/m_V$ increases. In Fig.~\ref{fig:Wj_pTspec} we show the $p_{T}$ distribution of the jet in the LHC environment both at LO and NLO, obtained using the program MCFM~\cite{Campbell:2002tg,Campbell:2003hd} using the same cuts as in Ref.~\cite{Campbell:2003hd}. One can clearly see that the discrepancy grows with increasing  $p_{T}$. Note that for $V+j$ final states the transverse momentum of the jet is equal to the transverse momentum of $V$, and the ratio of scales can therefore be constructed out of the kinematics of the $V$ boson alone. 

For processes with at least two jets in association with the produced vector boson, the individual transverse momenta of the jets are no longer related to the transverse momentum of the vector boson. Since the transverse momenta of the jets can balance each other, the $p_T$ of each jet can get large independent of the transverse momentum of the vector boson. However, the same arguments go through as for $V+j$ final states, and logarithmic dependence on the ratio $p_T/m_V$ is expected again.

At higher orders in perturbation theory more and more powers of the large logarithms can arise. In general, for any ratio of scales $r$ one finds up to two powers of logarithms for each power of $\alpha_s$, and for sufficiently large ratios of scales $r$ at least the leading logarithmic (LL) terms of the form  $\alpha_s^n \log ^{2n}r$ have to be resummed to all orders to obtain a reliable prediction. 

Effective theory methods naturally separate widely different energy scales from one another, and they are ideally suited to understand the corresponding logarithmic terms. Effective field theories are designed to reproduce an underlying theory at long distances, while short distance physics are captured in Wilson coefficients, which have to be adjusted order by order in perturbation theory to match the full theory at short distances. 

The power of effective field theory techniques is that for processes that involve several widely separated energy scales $\Lambda_1 \gg \Lambda_2 \gg \ldots \Lambda_n$ these scales can be removed one by one by matching onto successive effective theories, where each effective theory is valid below one of the scales $\Lambda_i$ in the problem. The Wilson coefficients in general depend on a renormalization scale $\mu_i$, and logarithms of the ratio $r_i = \mu_i / \Lambda_i$ will appear in the expressions. However, since each matching calculation only depends on the single scale $\Lambda_i$ (larger scales have been removed in a previous matching step, and the matching calculation is independent of the smaller scales), the choice $\mu_i \sim \Lambda_i$ minimizes these logarithmic terms. One can then use the renormalization group equations (RGEs) to sum the logarithms of ratios $\mu_{i+1}/\mu_i$ between the different matching scales. 

Since processes observed at high energy colliders  typically contain light particles with large energies, the appropriate effective field theory which reproduces all long distance physics of the standard model (SM) is soft-collinear effective theory (SCET)~\cite{Bauer:2000ew,Bauer:2000yr,Bauer:2001ct,Bauer:2001yt}. SCET contains collinear fields to describe the light, energetic particles, which can only interact with one another via the exchange of soft particles. For details on the construction and applications of SCET we refer the reader to the literature. 

\begin{figure}[t]
\centering
\includegraphics[width=0.8\columnwidth]{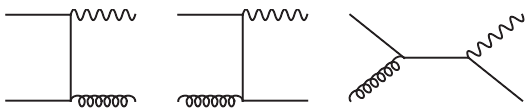}
\caption{Feynman diagrams contributing to the process $pp\to W^++j$.}
\label{fig:Wj_diagrams}
\end{figure}

We start by considering the process $pp \to V+j$, for which the tree-level amplitude can be calculated from the  diagrams shown in Fig.~\ref{fig:Wj_diagrams}. 
We try to separate the two scales $\Lambda_1 = p_{T}$ and $\Lambda_2 = m_V$ by using successive 
matching calculations. We will keep our discussion of the matching calculation very schematic, and ignore all complications, for example those due to operator mixing. This is sufficient to understand the resummation of the leading logarithmic terms at tree level, which is the aim of this letter. However,  the procedure can easily be generalized in a completely straightforward manner to allow the inclusion of higher order corrections.

\begin{figure}[t]
\centering
\includegraphics[width=0.65\columnwidth]{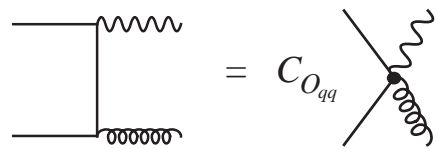}
\caption{Matching of the amplitude for the partonic subprocess $qq \to Vg$. A similar relation holds for $qg \to Vq$}
\label{fig:Wj_SCET}
\end{figure}

The intermediate propagator in Fig.~\ref{fig:Wj_diagrams} scale as $1/p_{T}^2$ for large values of $p_{T}^2$, and only gives rise to short distance physics. Thus, the same amplitude is reproduced in the effective theory using operators containing only the incoming partons and the outgoing parton and vector boson, and the effect of the intermediate propagator is contained in the Wilson coefficients $C_i$ of these operators, as is illustrated in Fig.~\ref{fig:Wj_SCET}.  Note that 
since we consider events with $p_{T} \gg m_V$, we can treat the vector boson as massless and as part of SCET. As argued above, the logarithms in the matching coefficients are minimized if one chooses $\mu_1 \sim \Lambda_1 = p_T$ in this calculation. Thus, we find for the amplitude
\begin{align}
A  \sim  C_{O_{qq}}(p_T) \langle {\cal O}_{qq} \rangle _{p_T}
+ C_{O_{qg}}(p_T) \langle {\cal O}_{qg} \rangle _{p_T}\,,
\end{align}
where $\langle {\cal O}_{qq} \rangle$ denotes the matrix element of the operator shown on the RHS of Fig.~\ref{fig:Wj_SCET}, and $ {\cal O}_{qg}$ denotes a similar operator for $qg$ initial states.
Note that the matrix elements of the operators $ {\cal O}_{ij}$ are evaluated at the scale $\mu = p_T$, as indicated by the subscript. The matrix element of the operator at a different scale can be obtained using the RGE of the effective theory, 
\begin{equation}
\langle {\cal O}_{ij} \rangle _{p_T} = U_{{\cal O}_{ij}}(p_T,\mu)\langle {\cal O}_{ij} \rangle _{\mu} \,,
\end{equation}
and the evolution kernels $U_{{\cal O}_{ij}}$ can be calculated straightforwardly in perturbation theory.

\begin{figure}[t]
\centering
\includegraphics[width=0.65\columnwidth]{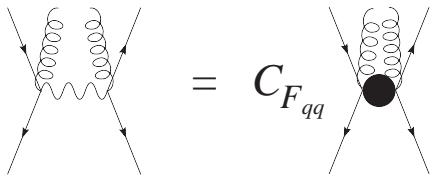}
\caption{Matching of the forward scattering matrix element for the partonic subprocess $qq \to Vg$. A similar relation holds again for $qg \to Vq$}
\label{fig:Wj_forward}
\end{figure}

The next scale to integrate out is the electroweak scale, and with it $m_V$. Since below this scale we can no longer treat the vector boson $V$ as a propagating particle in the effective theory, we have to remove $V$ from the theory altogether. This can be accomplished by considering the forward scattering matrix element as shown in Fig.~\ref{fig:Wj_forward} and 
integrating the intermediate $V$ boson out of the theory. The corresponding Wilson coefficients $C_{{\cal F}_{ij}}$ will depend on $m_V$, and the associated logarithms can be minimized by choosing $\mu_2 \sim \Lambda_2 = m_V$. Thus, the differential cross section is given by 
\begin{align}
\df \sigma  \sim & \, C_{{\cal O}_{qq}}^2(p_T) U_{{\cal O}_{qq}}^2(p_T,m_V) C_{{\cal F}_{qq}} (m_V) \langle {\cal F}_{qq} \rangle_{m_V}
\nn\\
& + C_{{\cal O}_{qg}}^2(p_T) U_{{\cal O}_{qq}}^2(p_T,m_V) C_{{\cal F}_{qg}}  (m_V)\langle {\cal F}_{qg} \rangle_{m_V}\,,
\end{align}
where $\langle {\cal F}_{qq} \rangle$ denotes the matrix element of the operator shown on the RHS of Fig.~\ref{fig:Wj_forward}, and $\langle {\cal F}_{qg} \rangle$ is the obvious generalization to $qg$ initial states. 

Finally, we have to calculate the matrix elements of the remaining operators. This can be accomplished using the factorization properties of QCD and SCET, such that the remaining long distance physics is absorbed into the parton distribution functions of the incoming partons and the jet functions of the outgoing partons. For sufficiently inclusive jet definitions, the jet properties can be calculated entirely in perturbation theory, and one writes
\begin{align}
\langle {\cal F}_{ij} \rangle_{m_V} &=  U_{{\cal F}_{ij}}(m_V,\mu) \langle {\cal F}_{ij} \rangle_{\mu} 
\nn\\
&=   U_{{\cal F}_{ij}}(m_V,\mu)  E_{ij}(\mu) f_i(\mu) f_j(\mu)\,.
\end{align}
The scale $\mu$ denotes the scale at which the parton distribution functions are evaluated, the Kernel $U_{{\cal F}_{ij}}(m_V,\mu)$ denotes the RG evolution of the operator from the scale $m_V$ to the scale $\mu$, and $E_{ij}$ denotes extra perturbative terms that depend on the scheme used for the parton distribution functions. 

Combing these results, and using the fact that by construction the combination of all matching coefficients reproduces the full partonic cross section expressions, the result reads
\begin{align}
\df \sigma  \sim & \sum_{ij} \df \hat\sigma_{ij}(x_i,x_j) f_i(x_i,\mu)f_j(x_j,\mu) 
\nn\\
& \times U_{{\cal O}_{ij}}^2(p_T,m_V) U_{{\cal F}_{ij}}(m_V,\mu) \,.
\label{eq:Wj_result1}
\end{align}
All logarithmic terms of the ratios $m_V/p_T$ and $\mu/m_V$ have been resummed in the kernels  $U_{{\cal O}_{ij}}(p_T,m_V)$ and $U_{{\cal F}_{ij}}(m_V,\mu)$. However, the result in \eqref{eq:Wj_result1} can be simplified further by using a simple relation between the two evolution kernels, which is correct at LL accuracy. The leading logarithmic structure is governed by the so-called cusp anomalous dimension of the operators ${\cal O}_{ij}$ and ${\cal F}_{ij}$, which can be calculated in SCET from collinear one-loop diagrams involving only strongly interacting collinear fields in a given direction. Since the operator ${\cal F}_{ij}$ is obtained from the forward matrix element of two operators ${\cal O}_{ij}$ by integrating out the vector boson which is not strongly interacting, the evolution kernels are related at LL order by the simple relation
\begin{equation}
U^{\rm LL}_{{\cal F}_{ij}}(\mu_1,\mu_2) = \left[U^{\rm LL}_{{\cal O}_{ij}}(\mu_1,\mu_2)\right]^2 \,.
\label{eq:simple_relation}
\end{equation}
Since furthermore the evolution kernels satisfy the simple relation $U(\mu_1,\mu_2)U(\mu_2,\mu_3) = U(\mu_1,\mu_3)$, we can simplify our result as
\begin{align}
\df \sigma^{\rm LL}  \sim & \sum_{ij} \df \hat\sigma_{ij}(x_i,x_j) f_i(x_i,\mu)f_j(x_j,\mu) 
U^{\rm LL}_{{\cal F}_{ij}}(p_T,\mu) \,.
\label{eq:Wj_result2}
\end{align}
Thus, by choosing to evaluate the parton distribution functions at a dynamical scale $\mu = p_T$, the dependence on the evolution kernel disappears entirely, and for that scale choice the result is just given by the usual fixed order expression. 

This result is not unexpected, since it simply states that the relevant scale for events with $p_T \gg m_V$ is given by the large transverse momentum in the process, rather than the mass of the vector boson. In order to cover processes for which $p_T \ll m_V$ (complementary to our previous assumption), for which the relevant scale should be $\mu \sim m_V$, several analyses have used the dynamical scale $\mu = [m_{V}^2 + p_{T_V}^2]^{1/2}$ \cite{NLOWjjjBlackhat,Aaltonen:2007ip,Kuhn:2004em,Stump:2003yu}. Since for $V+j$ final states the transverse momentum of the vector boson $p_{T_V}$ is equal to that of the jet, our result shows that this scale setting does indeed sum the large logarithms of the form $\log m_V/p_T$.

\begin{figure}[t]
\centering
\includegraphics[width=0.8\columnwidth]{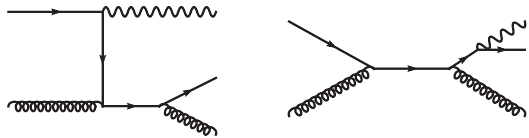}
\caption{Sample Feynman diagrams for initial-state radiation and final-state radiation contributions to $pp \to V + 2$ jets.}
\label{fig:Wjjdiagrams}
\end{figure}

\begin{figure}[t]
\centering
\includegraphics[width=1.0\columnwidth]{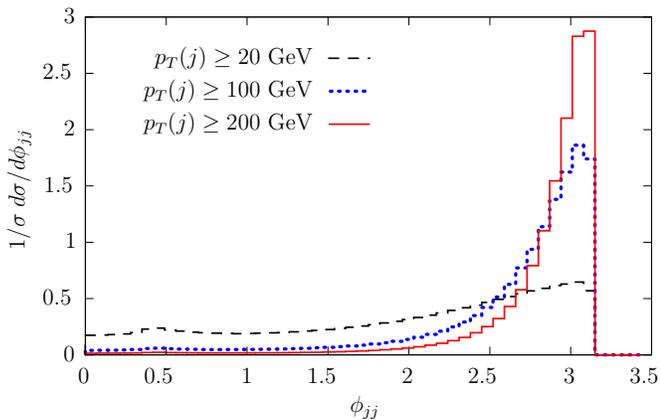}
\caption{Distribution in $\phi_{jj}$ for $pp\to W^++2j$ and for different cuts on the minimal jet transverse momenta.}
\label{fig:phiPlot}
\end{figure}

For processes with two or more jets in the final state, large logarithmic terms will originate from phase space integrations. However, these logarithms can still be resummed using a very similar analysis as performed for $pp \to V+j$ case. To show this we study the kinematics for $pp \to V+2j$. Many Feynman diagrams contribute to this process, but for large partonic center of mass energy the dominant contribution to the cross section arises from diagrams where the $V$ boson is radiated from an initial or final quark with either soft momentum or collinear to the quark it is radiated from. This gives rise to effects that are enhanced by a factor of $p_T/m_V$. Two representative Feynman diagrams of such initial or final state radiation are shown in Fig.~\ref{fig:Wjjdiagrams}. Both of these processes result in kinematics where the two jets are back-to-back in the transverse plane, with the angle $\phi_{jj}$ between them near $\pi$. We illustrate this effect in Fig.~\ref{fig:phiPlot}, where the distribution in $\phi_{jj}$ for different values of a cut on the smallest available $p_T$ of the jets is shown. One can see that the larger the cut on $p_T$, the more the distributions peak towards $\phi_{jj} = \pi$, in agreement with our expectations. Note that there is another kinematical region that is enhanced by a large propagator, which is when the two jets are collinear to each other, and are recoiling against the $V$ boson. It is the size of the jets which regulates this singularity, and in fact the small feature near $\phi_{jj}=0.4$ is due to this kinematical configuration ($\Delta R = 0.4$ was used in the jet algorithm). However, this region is enhanced by only a power of $1/(\Delta R)$, and for large enough values of $p_T$ the initial and final state radiation of the $V$ boson dominates. 

\begin{figure}[t]
\centering
\includegraphics[width=0.8\columnwidth]{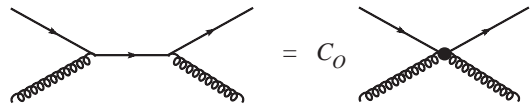}
\caption{First matching step at the high scale $\mu \sim p_T$.}
\label{fig:Wjjmatch}
\end{figure}

To resum the logarithms of order $m_V/p_T$, we again perform a two step matching calculation. In the first step, we integrate out the propagators with virtuality of order $p_T^2$, resulting in a matching calculation as indicated in Fig.~\ref{fig:Wjjmatch}. In order to avoid large logarithmic terms in this matching calculation, we choose $\mu \sim p_T$. Note that for the dominant initial and final state radiation contributions the $V$ boson does not participate in this matching calculation. In a second matching step, the $V$ boson, together with all other propagators with virtuality of order $m_V$ are integrated out. The natural scale for this matching is again $\mu \sim m_V$. Finally, the matrix elements are calculated, resulting in parton distribution functions evaluated at that final scale $\mu$. In order to connect the different scales to one another, RG evolution is used, and using a similar argument as in the case of $V+j$ final states, one can show that to LL order the evolution kernel $U_{\cal O}$ connecting the scales $p_T$ to $m_V$ is related to the kernel $U_{\cal F}$ describing the running from $m_V$ to $\mu$ by \eqref{eq:simple_relation}. Thus, all leading logarithms can again be summed by choosing the factorization scale to be of order $p_T$ of the jets. Note that despite the derivation being so similar as for the case of $V$ plus a single jet, to our knowledge the scale choice derived in this work has not been used in any published results\footnote{An observations that the traditional scale choice in LO calculations gives poor convergence with NLO calculations has recently been made~\cite{DixonTalk}, together with a suggestion for using a dynamical scale similar to ours in spirit. For vector boson fusion, dynamical scales have also been found empirically to improve the convergence of perturbation theory~\cite{Bozzi:2007ur}.}.

Unlike the change from $n=1$ to $n=2$ in $pp\to V+n$ jets processes, a further increase to $n\ge 3$ does not give rise to qualitatively new kinematic situations, in that individual transverse jet energies can far surpass the vector boson's transverse energy. For any given set of outgoing 4-momenta, there are of course more scales in the problem, but for generic events, where each of the jet energies scales like their respective $p_T$, the very same framework based on EFTs can be applied. 

\begin{figure*}[t]
\centering
\includegraphics[width=1.0\columnwidth]{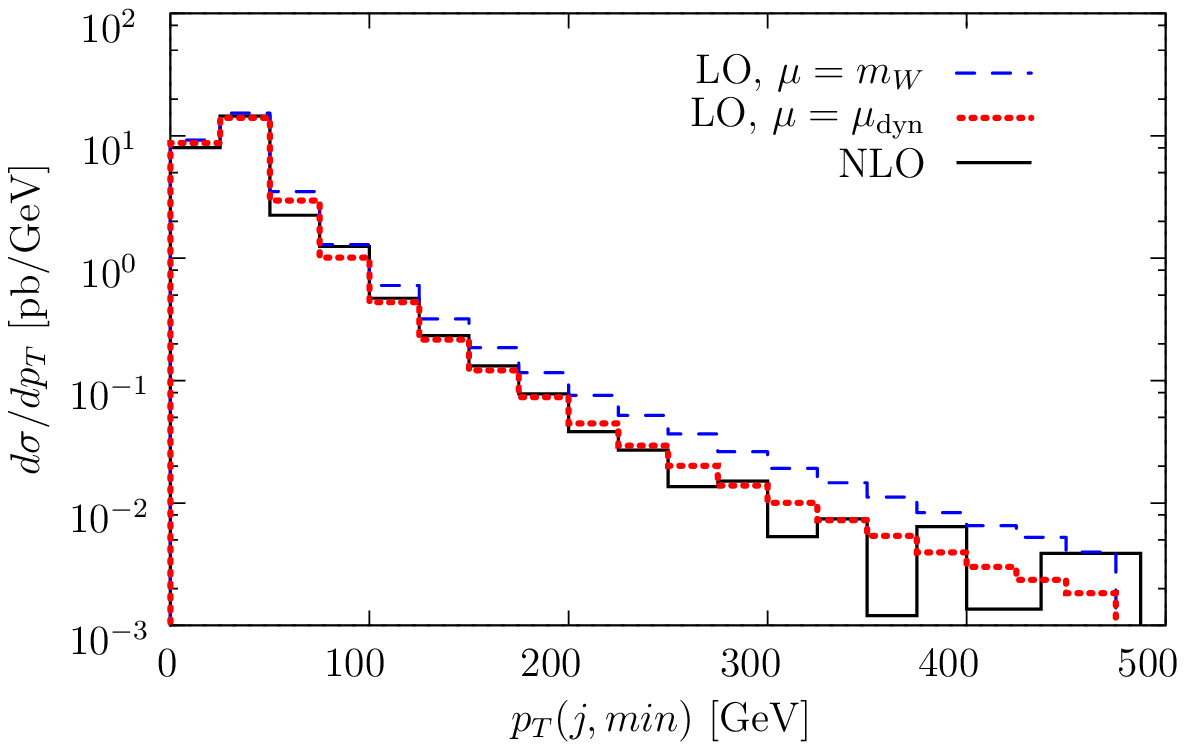}
\includegraphics[width=1.0\columnwidth]{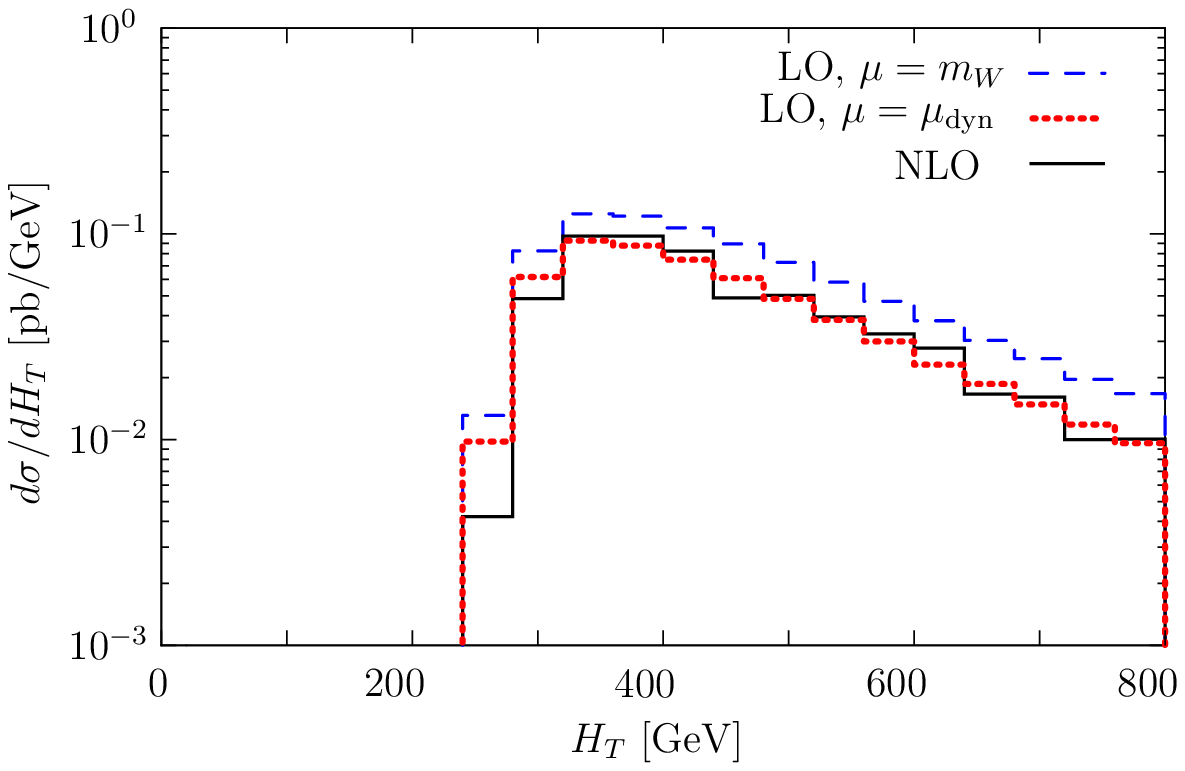}
\caption{Comparison between the two scale settings $\mu=m_V$ (dashed, blue) and $\mu=\mu_{\rm dyn}$ (dotted, red) and the NLO prediction (solid, black) for the process $pp \to W^++2j$. Left: $p_T$ spectrum of the second jet with a cut of $p_T \ge 20$ GeV. Right: $H_T$ spectrum with $p_T \ge 100$ GeV . For high-$p_T$ events the NLO calculation becomes less reliable as statistical fluctuations gain more prominence.}\label{fig:Scale_comp}
\end{figure*}

We now study the effect of the resummation of the leading logarithmic terms by comparing the LO calculation with the new choice of scale $\mu \sim p_T$ with the LO calculation using the traditional scale choice $\mu \sim m_V$ and the NLO calculation with $\mu=m_V$. Note that our scale setting procedure only resums the logarithms correctly at LO/LL accuracy, and we therefore do not change the scale used in the NLO calculation. To be concrete, we chose\footnote{This scale choice is somewhat ad hoc, but consistent with its scaling with $p_T$. Other choices which satisfy the correct scaling with $p_T$ differ at most by logarithms which are not enhanced in the limit $m_V/p_T \to 0$. } $\mu_{\rm dyn}^2 = (m_{\rm hadr}/2)^2+m_V^2$, where $m_{\rm hadr}$ is the dijet invariant mass of the hadronic system. The scale dependence at NLO is considerably reduced over the dependence at LO, such that the choice of scale used in the NLO calculation does not affect our conclusions. We will present results for $W$ + jets in the final state, but we have checked that our results apply equally well for $Z$ + jets in the final state. 

In Fig.~\ref{fig:Scale_comp} we show on the left the distribution of the transverse momentum of the softer jet in $pp \to W^++2j$ events. For small values of $p_T$ the two different choices for the scale give comparable results, with both results agreeing reasonably well with the NLO calculation. For larger values of $p_T$ the LO calculation with the traditional choice of scale setting $\mu = m_V$ disagrees more and more with the NLO result, while the LO calculation with the dynamical scale setting agrees with the NLO results over the full range of $p_T$. To illustrate that this is not just an artifact of the distribution in transverse momenta, we also show the $H_T$ distribution, where we have imposed a cut $p_T > 100$ GeV to select events where the difference between the two scale setting choices becomes more pronounced. Again, we see that that the LO calculation for the traditional choice of scale setting $\mu = m_V$ does not agree well with the NLO calculation, but that the new scale choice improves the situation significantly. We have compared the traditional and new scale setting for many other observables, and in all cases the new choice of scale improves the distributions.

\begin{figure*}[t]
\centering
\includegraphics[width=1.0\columnwidth]{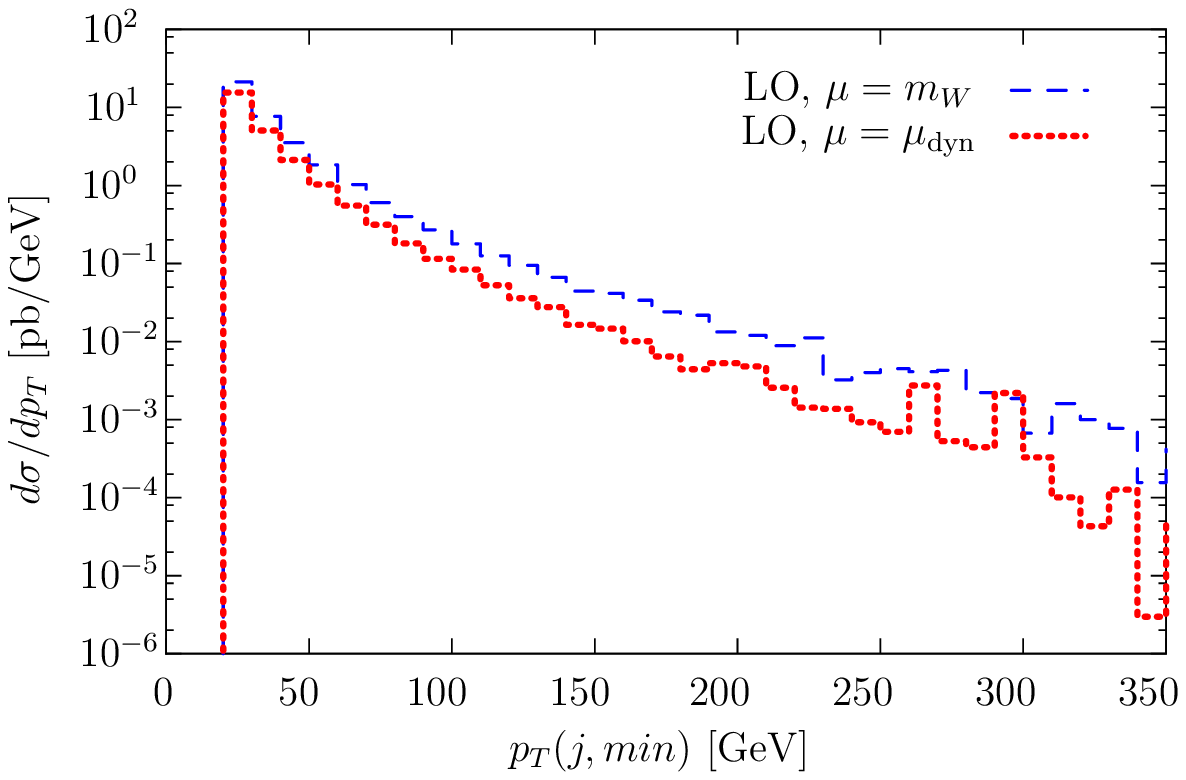}
\includegraphics[width=1.0\columnwidth]{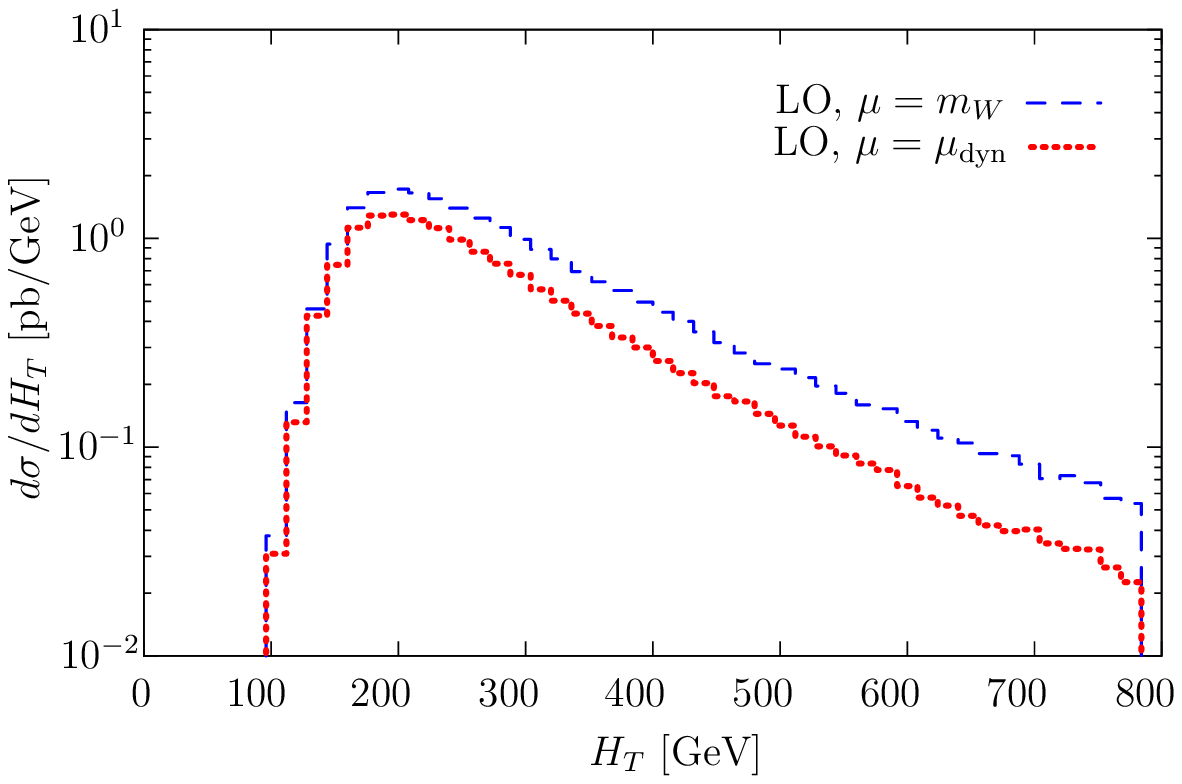}
\caption{Comparison between the two scale settings $\mu=m_V$ (dashed, blue) and $\mu=\mu_{\rm dyn}$ (dotted, red) for the process $pp \to W^++3j$. Left: $p_T$ spectrum of the third jet, Right: $H_T$ spectrum, both with a cut of $p_T \ge 20$ GeV.}\label{fig:jjjPlot}
\end{figure*}

In Figure~\ref{fig:jjjPlot} we compare for $n=3$ the LO predictions for the $p_T(j,min)$ spectrum -- the smallest of the three transverse jet momenta -- with traditional scale setting and with the dynamical scale setting. We again observe that the net effect of resummation is a lowering of the differential cross section, and that the region of large $p_T$ receives a larger relative correction than the small $p_T$ region, i.e.~a correction of the shape of differential cross section. This agrees qualitatively with the heroic fixed-order NLO calculation recently performed \cite{Ellis:2009zw,NLOWjjjBlackhat}, both for the $p_T$ and similarly for the $H_T$ distributions.

It should be noted that the simple scale setting procedure used above gives results at leading order in perturbation theory with the leading logs resummed. The effective field theory methods introduced, however, can be used to improve these results by including higher orders in fixed order perturbation theory and/or higher order terms in the resummation calculation. NLO corrections, for example, can be included by calculating the matching coefficients $C_{{\cal O}_{ij}}$, $C_{{\cal F}_{ij}}$ and $E_{ij}$ to higher orders in perturbation theory, while higher orders in the resummation can be included by computing the evolution kernels $U_{{\cal O}_{ij}}$ and $U_{{\cal F}_{ij}}$ to higher orders. Note that we expect the relation between the two evolution kernels given in \eqref{eq:simple_relation} to be violated at NLL order, such that the NLL resummation can not be accomplished by a simple choice of scales in the parton distribution functions.

In this letter we have investigated the effect of leading-logarithmic resummation for jet production with associated vector boson ($V=W,Z$) production at the LHC. The resummation is achieved via a series of matching onto versions of soft-collinear effective theory and running of the coefficient functions to their natural scales using their renormalization-group properties. The result simplifies to the known partonic cross section expression folded with the appropriate parton distribution functions if the parton distribution functions are evaluated at a high scale which scales like the transverse momentum of the jets. The main purpose of resummation is to improve the convergence of the perturbative expansion which we have demonstrated specifically for events with high transverse momenta in $pp \to V+n$ jets processes.

\acknowledgments

We are grateful to Jesse Thaler, Beate Heinemann and Lance Dixon for helpful conversations. 
This work was supported in part by the Director, Office of Science, Offices
of High Energy and Nuclear Physics of the U.S.\ Department of Energy under the
Contracts DE-AC02-05CH11231, and an LDRD from 
Lawrence Berkeley National Laboratory.

\end{document}